\documentclass[pdflatex,sn-mathphys-num]{sn-jnl}

\usepackage{graphicx}%
\usepackage{multirow}%
\usepackage{amsmath,amssymb,amsfonts}%
\usepackage{amsthm}%
\usepackage{mathrsfs}%
\usepackage[title]{appendix}%
\usepackage{xcolor}%
\usepackage{textcomp}%
\usepackage{manyfoot}%
\usepackage{booktabs}%
\usepackage{algorithm}%
\usepackage{algorithmicx}%
\usepackage{algpseudocode}%
\usepackage{listings}%
\usepackage{lmodern}

\begin{document}

\title[Article Title]{Decoding the city: multiscale spatial information of urban income}


\author*[1,2,3]{\fnm{Lu\'is M. A.} \sur{Bettencourt}}\email{bettencourt@uchicago.edu}

\author[1]{\fnm{Ivanna} \sur{Rodriguez}}

\author[1,4]{\fnm{Jordan T.} \sur{Kemp}}

\author[5]{\fnm{Jos\'e} \sur{Lobo}}

\affil*[1]{\orgdiv{Dept. of Ecology and Evolution, Dept. Sociology}, \orgname{University of Chicago}, \orgaddress{ \city{Chicago}, \postcode{60637}, \state{IL}, \country{USA}}}

\affil[2]{ \orgname{Santa Fe Institute}, \orgaddress{ \city{Santa Fe}, \postcode{87501}, \state{NM}, \country{USA}}}

\affil[3]{ \orgname{Vienna Complexity Science Hub}, \orgaddress{ \city{Vienna}, \postcode{1030}, \country{Austria}}}

\affil[4]{\orgdiv{Institute for New Economic Thinking at the Martin School}, \orgname{University of Oxford}, \orgaddress{ \city{Oxford }, \postcode{OX1 3UQ}, \country{UK}}}

\affil[5]{\orgdiv{School of Sustainability, College of Global Futures}, \orgname{Arizona State University}, \orgaddress{ \city{Tempe}, \postcode{85281}, \state{AZ}, \country{USA}}}


\abstract{Cities are characterized by the coexistence of general aggregate patterns, along with many local variations. This poses challenges for analyses of urban phenomena, which tend to be either too aggregated or too local, depending on the disciplinary approach. Here, we use methods from statistical learning theory to develop a general methodology for quantifying how much information is encoded in the spatial structure of cities at different scales. We illustrate the approach via the multiscale analysis of income distributions in over 900 US metropolitan areas. By treating the formation of diverse neighborhood structures as a process of spatial selection, we quantify the complexity of explanation needed to account for personal income heterogeneity observed across all US urban areas and each of their neighborhoods. We find that spatial selection is strongly dependent on income levels with richer and poorer households appearing spatially more segregated than middle-income groups. We also find that different neighborhoods present different degrees of income specificity and inequality, motivating analysis and theory beyond averages. Our findings emphasize the importance of multiscalar statistical methods that both coarse-grain and fine-grain data to bridge local to global theories of cities and other complex systems.}




\maketitle

\section{Introduction}\label{sec1}
Complex systems – such as cities, ecosystems, or organisms - are often recognizable by the coexistence of general structures along with many local variations~\cite{goldenfeld_simple_1999,Bettencourt2021}. Such a description is deceptively simple though, because it glosses over the fact that local variations in a patch of forest or a city street represent not just random fluctuations~\cite{Bettencourt2010,bettencourt_towards_2019}, as in physical systems, but a long history of serendipity and adaptation emphasized in biology and the social sciences~\cite{crow_introduction_1970,Sampson2012}. This distinction is central to any theoretical approach to cities: How to account for mechanisms of change and adaptation not only on large scales, but also locally in neighborhoods or small groups? What is chance and what is necessity in different places, at different times? How may local learning and adaptation lead to large-scale growth and development~\cite{ jones_new_2010,Kemp2024}?

It is rather self-evident that the aggregate (or macro level) characteristics of an urban area should emerge from the interactions occurring within~\cite{jacobs_death_2011,Batty2008,Bettencourt2013}. Recovering what these interactions are and identifying how they generate the macroscopic behavior is confused by the well-known “ecological fallacy”~\cite{Robinson1950}. This issue results from inferences about the nature of individuals made on the basis of group averaged characteristics~\cite{Robinson1950,Bettencourt2021}:  Aggregating and averaging data invariably masks underlying heterogeneity and, in complex systems, may destroy relevant information. 

The present paper develops a systematic approach to the analysis of how individual agent characteristics (such as income and other socioeconomic and demographic variables) form different patterns of organization at different scales. We use data from US metropolitan areas to motivate and illustrate our results, bridging large-scale patterns of income distribution at the metropolitan area scale (city) to those observed in small areas, such as block groups (neighborhoods), which are often quite different from the city at large. 

The conceptual and empirical obstacles to building theories of complex systems across different scales are evident from the diverging approaches developed by different disciplines~\cite{anderson_more_1972,jacobs_death_2011,Batty2008,Bettencourt2010}. For example, when modeling cities, physicists and economists tend to prefer the study of averaged behavior and adopt (homogeneous) representative agents~\cite{AxtellFarmer2025}, while other social scientists emphasize the specificity of places and people.  Thus, from a statistical perspective, different approaches can be roughly divided into two camps. One emphasizes more aggregate data and modeling, while the other claims that such aggregation is (sometimes) a gross approximation to relevant phenomena and prefers local evidence and more attention to context and history. Here, our goal is not to adjudicate between these approaches but to demonstrate explicitly that both are necessary for the study of complex systems. In particular, we show that the complexity of explanation at different scales in cities can be quantified and pinpointed by appropriate methods of information theory.

Our approach contrasts with the common strategy in statistical physics --known as coarse-graining-- of obtaining predictable macroscopic statistics from averaging over smaller and faster scales~\cite{goldenfeld_lectures_1992}. Coarse-graining is appropriate when variations on small and faster scales are random, resulting from microscopic disorder. Such approach then tells us that, in many known systems, most local details do not contribute to macroscopic behavior.  This is the basis for the renormalization group in statistical physics, which constitutes the essential tool to analyze phase-transitions in bulk materials and has resulted in powerful ideas of universality~\cite{kadanoff_statistical_2000,anderson_more_1972}. 

It follows that proceeding in the direction of coarse-graining leads to information loss as local states are replaced by averages over larger scales~\cite{kadanoff_statistical_2000}, see Supplementary Text 1.1 for derivation. Because of this essential feature, renormalization group methods are not invertible. By contrast, proceeding in the opposite direction, from more averaged to less averaged systems -- which we call “fine-graining”-- requires the addition of information as new degrees of freedom on smaller scales must be specified. Such “fine-graining” methods have now been developed in statistical learning theory and inference~\cite{mackay_information_2003,SRCNN2014,SRGAN2017}, and in evolutionary biology in terms of selection~\cite{DonaldsonMatasci2010,BettencourtGrandisonKemp2025}. Following these developments, we see great opportunities for the use of emerging generative models from artificial intelligence to account for the detailed statistical structure of cities. Here, we lay some basic foundations for such future efforts by demonstrating and discussing some of the structures actually observed in US cities.    

In pursuing an information-theoretic approach, we are  also motivated by accounting in a new way for the complexity of human behavior in cities, and specifically the observed patterns of spatial sorting of households into neighborhoods by personal income~\cite{Sampson2012,Wilson1987,Bruch2014,Intrator2016}. The differential sorting of households into neighborhoods, whether by income, ethnicity or any other characteristic, is a classic problem in sociology and economics~\cite{park_city:_2010,schelling_dynamic_1971,Wilson1987,Sampson2012}, approached originally by showing that seemingly innocuous local decisions can lead to extreme macroscopic patterns of segregation~\cite{schelling_dynamic_1971}. We will show, however, that observed patterns of economic sorting in US metropolitan areas are, generally, place and income group specific. This specifies how they entail more complex choices and therefore require different decision models to explain them in practice~\cite{BruchMare2006}. 

\section{Results}\label{sec2}
\begin{figure}[ht]
\centering
\includegraphics[width=0.9\textwidth]{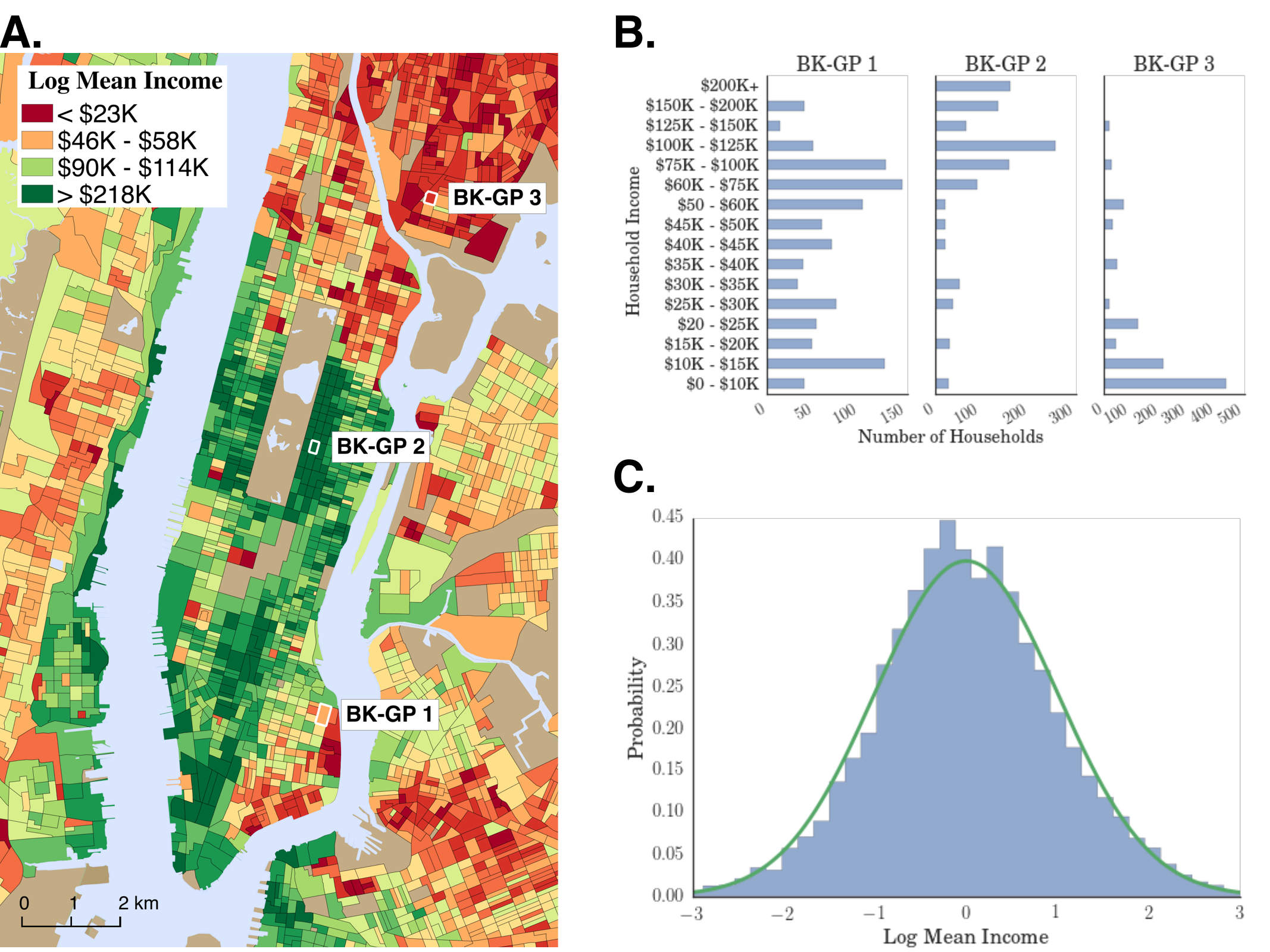}
\caption{The heterogeneity of neighborhoods in New York City. A.  Average household income in New York City census block groups. B. Income distributions in selected neighborhood shown in Fig. 1A. C. The city-wide distribution of household income is well described by a lognormal distribution (green line), which we show in SOM is a very good general model for the household income distribution in all US MSAs. The mean and variance of this distribution obey simple scaling relations, see Supplementary Materials {\bf S2.2}. Data was compiled by the US Census 5-year American Community Survey (2010 release) comprising of over 200,000 block groups nationwide and about 14,000 in the New York Metropolitan Statistical Area (MSA).}\label{fig1}
\end{figure}

To illustrate our methods and objectives, consider the pattern of household income in New York City neighborhoods (Figure \ref{fig1}A); see Figs. S1-S5 for details and other cities. 
We observe strong heterogeneity at different spatial scales, from adjacent neighborhoods (delineated by block groups, BK) with different average household incomes to larger patches of wealth and poverty, e.g. the Upper East Side (the richest part of Manhattan) or the Bronx (generally poor). Moreover, we observe  different income distributions inside each neighborhood, Fig.~\ref{fig1}B, so that poor neighborhoods contain people who are not poor (BK3), and rich neighborhoods many who are not rich (BK2). As is well known, this spatial heterogeneity is long-lived, persisting for decades or longer~\cite{Sampson2012}, through many economic cycles and substantial demographic turnover. 

Interestingly, such rich and detailed patterns contrast with the simple normal distribution for (the logarithm of) household income across the entire city (metropolitan area), Fig. \ref{fig1}C. A log-normal process emerges from general multiplicative growth processes, where each contributing factor operates independently. In this case wealth appreciates at a fluctuating rate~\cite{gabaix_zipfs_1999,kemp_statistical_2022}. This coarse-grained statistic is common to all US metropolitan areas: the distribution of income across all cities is well described by a lognormal (except at the top tail, which is censored in this data), see Supplementary Text S1.2 and Figures S6-7. Moreover, the two parameters characterizing this distribution are themselves simple and general, see Supplementary Text 1.3 for background: The mean obeys a scaling relation~\cite{Bettencourt2007,GomezLievano2012}, well parameterized by a power-law $\langle Y(N,t))\rangle = Y_0 (t) N^\beta(t)$, see Fig. S8,
with general system-wide parameters $Y_0(t)$, $\beta$~\cite{Bettencourt2021}; where the exponent $\beta>1$, expresses urban superlinear (agglomeration) effects~\cite{Bettencourt2007,bettencourt_hypothesis_2013}, Fig. S8. Its value is predictable from urban scaling theory, which describes the city in terms of interdependent networks of people, organizations and infrastructure~\cite{Bettencourt2013,Bettencourt2021}.  The variance of the logarithmic income, see  Fig. S9, is also a simple general number, associated with the cumulative volatility of incomes over time and setting the level of inequality within the urban area, as measured, for example, by the Gini coefficient or Theil’s $T$ index~\cite{kemp_statistical_2022}, Supplementary Text 1.4. 

A statistical regularity thus emerges, Fig. \ref{fig1}C, at the city-wide scale as the result of averaging over a rich pattern of local neighborhood (and individual) variations, Fig. \ref{fig1}A-B. While this coarse-grained statistic reveals important aspects of urban socioeconomic dynamics~\cite{schlapfer_scaling_2014,bettencourt_urban_2020}, we now focus on the structure of neighborhood variations using this macroscopic regularity as a reference. Specifically, we quantify the complexity of the pattern of variations at the neighborhood level by comparing income probability distributions at different levels of spatial aggregation, BK distributions in Fig.\ref{fig1}B to  Fig.\ref{fig1}C.  To do this explicitly, we write
\begin{eqnarray}
p(y_\ell|n_j)= w_{\ell j} p(y_\ell)            
\label{eq:1}
\end{eqnarray}
where $p( y_\ell|n_j)$  
is the distribution (normalized share) of income $y$, in discrete bins labeled by $\ell$ in neighborhood $n_j$ (the different colorful BKs in Fig \ref{fig1}A). Here, 
$p(y_\ell)$ is the income distribution at a more aggregate level (Fig. \ref{fig1}C), which we take to be the entire city (metropolitan area) throughout this paper. Eq. \ref{eq:1} defines the weights, $w_{\ell j} \equiv p(y_\ell|n_j)/p(y_\ell)$ which transform one distribution into the other, see Fig. S10.  With this definition, the average weights over income obey the normalization $\langle w_j \rangle= \sum_\ell w_{\ell j} p(y_\ell)=1$ for all neighborhoods $j$, see Methods. 

Eq.~\ref{eq:1} is well known and can be readily recognized from two different perspectives. First, it is the haploid model of population genetics~\cite{crow_introduction_1970,frank_natural_2011}, also known as the ”replicator” equation in evolutionary game theory~\cite{PageNowak2002}. In that context, the two distributions are related across time (not space) and the weights $w_{\ell j}$ are the relative fitness of a trait $\ell$, expressing its differential propagation to the next generation. The stronger the deviation of $w_{\ell j}$  away from the average (unity), the stronger the selection for allele $\ell$. This corresponds to high fitness if $w_{\ell j}>1$, and vice-versa if $w_{\ell j}<1$. When $w_{\ell j}=1$, the dynamics is neutral, and there is no selection. Selection refers to the process by which certain elements are favored over others based on specific features of the environment. The replicator equation allows the fitness function to incorporate the distribution of the population types rather than setting the fitness of a particular type constant, thus capturing the essence of selection~\cite{BettencourtGrandisonKemp2025}. This interpretation gives a mathematical correspondence between evolutionary dynamics (in time) and neighborhood sorting (in space).   

Second, Eq.~\ref{eq:1} is a form of Bayes’ theorem~\cite{mackay_information_2003}, which leads to the interpretation of $w_{\ell j}$  in terms of probability ratios, specifically  
\begin{eqnarray}
p(y_\ell|n_j)=\frac{p(n_j|y_\ell)}{p(n_j)} p(y_\ell) \quad \rightarrow \quad
w_{\ell j }= \frac{p(n_j|y_\ell)}{p(n_j)} = \frac{p(y_\ell|n_j)}{p(y_\ell)}
=\frac{p(y_\ell,n_j)}{p(y_\ell ) p(n_j)}.
\label{eq:2}
\end{eqnarray}
Here, $p(n_j |y_\ell)$ is the probability for a person in the city to reside in neighborhood $n_j$, given that they have income $y_\ell$. The probability, $p(n_j )$,  is the (income independent) probability to live in neighborhood $n_j$, estimated as the ratio of its population to that of the city, see Methods. 

This second perspective leads to another powerful correspondence between probability theory, inference, and neighborhood structure. In this context, $\log w_{\ell j}$, in Eq.~\ref{eq:2} is the (non-averaged) mutual information~\cite{mackay_information_2003} between neighborhood $j$ and the distribution of income $y$. (Mutual information measures how much knowing one random variable reduces uncertainty about another, quantifying the amount of shared information between them.) To see this more explicitly consider the average of $\log w_{\ell j} $ over income groups 
\begin{eqnarray}
\langle \log w_{ j} \rangle = \sum_l p(y_\ell|n_j) \log  \frac{p(y_\ell|n_j)}{p(y_\ell)}= D [ p(y|n_j ||p(y)].  
\label{eq:3}
\end{eqnarray}
Here, $D[...]$ is the  (Kullback-Leibler) divergence between the distributions of income city-wide and in neighborhood $j$. The Kullback-Leibler divergence is a core concept in information theory that quantifies how much one probability distribution differs from another, expected distribution—effectively measuring the information lost when the latter is used to approximate the former \cite{CoverThomas2006}. For each neighborhood $j$, it represents the amount of information required to describe its specific income distribution, assuming we begin with knowledge of the city-wide income pattern.  Atypical neighborhoods, with income distributions very different from the city as a whole, require a longer explanation (more information). Neighborhoods that already reflect the city-wide pattern require no further description. In other words, atypical neighborhoods necessitate the invocation of local neighborhood effects, in addition to the city-wide distribution of traits common to all places, to explain their income distributions. The term $\langle \log w_j \rangle$ expresses the strength of neighborhood effects in each $j$ relative to city-wide dynamics, measured in units of information. 

\begin{figure}[ht]
\centering
\includegraphics[width=0.9\textwidth]{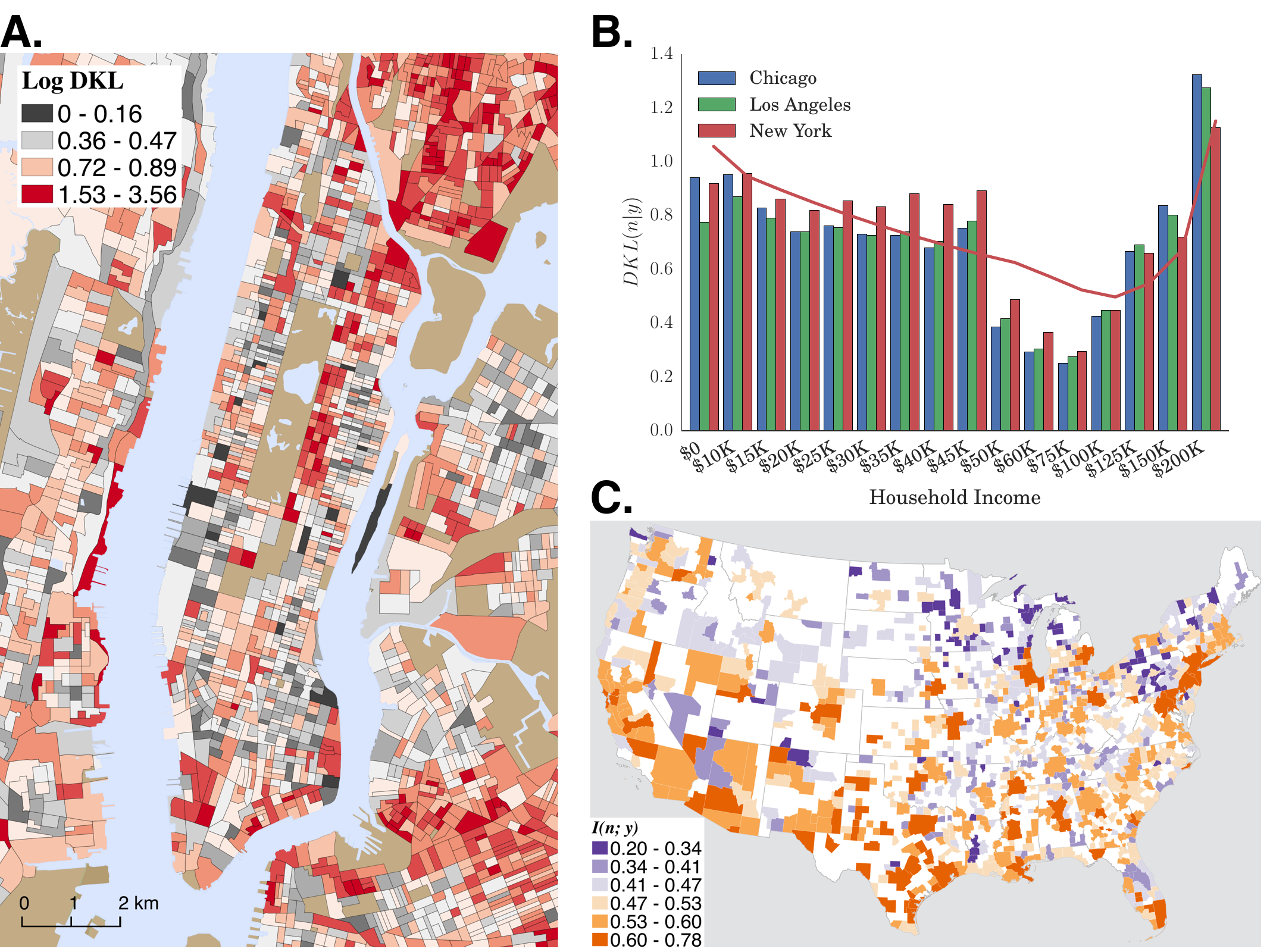}
\caption{
Spatial Selection Measured in Units of Information. A. The information, $\langle \log w_j \rangle$, necessary to explain the household income distribution of different neighborhoods,  given the city-wide income statistics, Fig. 1C.  The entropy of the income distribution for the New York Metropolitan Area is $H(y)=3.89$ bits. Thus neighborhoods in darker gray require very little additional information, while those in red may demand local explanations that are as comprehensive as the city-wide pattern of income itself. Comparing to Fig. 1A, note that it is both the poorest and richest neighborhoods that tend to require more information (red).  B. The average intensity of local selection by income, measured by $\langle \log w_l \rangle $, for several large US MSAs. The richest income brackets experience the strongest spatial selection, followed by the lowest incomes. The variation with income can be reasonably fitted by a quadratic form (red line, for New York MSA) with $\langle \log w_l \rangle = 1.058 -1.167 10^{-5} y+ 6.074 x 10^{-11} y^2$. C. The average strength of neighborhood effects across urban areas in the US measured by the mutual information $I(y;n)$  (942 metropolitan and micropolitan areas).  Orange denotes stronger average neighborhood selection, where the distribution of income in each of the city's neighborhoods is less like that of the city as a whole, and vice versa (purple).  Cities with low $I(y;n) \rightarrow 0$ have less distinguishable neighborhood structure by income.
}\label{fig2}
\end{figure}

Fig.~\ref{fig2}A shows the strength of income neighborhood effects for each block-group in New York City, measured by $\langle \log w_j \rangle$, see Figs.~ S11-S15 for details and other cities. We observe a very mixed pattern of local selection with many neighborhoods reflecting the distribution of income for the city as a whole (dark gray). However, others display a strong local flavor (red). We verified that the magnitude of the observed differences could not be the result of a purely random process of drawing individuals at random from the metropolitan income distribution into each place, Fig. S16.

Comparing Figs. \ref{fig1}A and \ref{fig2}A suggests that the most atypical neighborhoods have both the highest and the lowest average household incomes. It turns out that this is a general pattern of selection across all US metropolitan areas that we can quantify systematically via the average of $\log w_{\ell j}$ over neighborhoods $j$, 
\begin{eqnarray}
\langle \log w_\ell \rangle  = \sum_j p(n_j|y_\ell ) \log \frac{p(n_j|y_\ell)}{p(n_j)} = D [ p(n |y_\ell) || p(n)].    
\label{eq:4}
\end{eqnarray}
This quantity is the average information necessary to explain the distribution of specific income groups $y_\ell$ across the city, given that we know its neighborhood structure. In the absence of neighborhood effects (i.e., of spatial sorting), this quantity is zero, meaning that each level of income is distributed at random over space. Thus, the magnitude of Eq. \ref{eq:4} quantifies the differential average strength of neighborhood effects for different income levels in each city. Fig. \ref{fig2}B shows that the neighborhood effects are strongest for the highest income group, followed by the lowest. Middle-incomes are observed to be spatially the most mixed and thus less determined by specific neighborhoods. This is an interesting finding because it shows that different income groups exercise different kinds of choices – by preference and necessity - in terms of residential location. This also demonstrates that any realistic model of residential choice in US cities needs to be an explicit function of income levels.

These two effects are summarized in turn by a single quantity that captures the overall strength of neighborhood effects for each city in units of information, Fig. \ref{fig2}C. This is the total (mutual) information, $I(y;n)= \langle \log w \rangle $,
between neighborhood structure and income, given as the average of the previous quantities over the remaining variable, 
\begin{eqnarray}
\langle \log w \rangle &&= \sum_j p(n_j) D [p(y|n_j) || p(y) ] = \sum_\ell p(y_\ell) D [p(n|y_\ell ) || p(n)] \\ 
&&= \sum_{\ell,j} p(y_\ell, n_j ) \log w_{\ell j} = I(y;n). \nonumber   
\end{eqnarray}
If every neighborhood were a microcosm of the city as a whole, then all income groups would be spatially well-mixed and there would be no (income) neighborhood effects, leading to $I(y;n)=0$. Conversely, in cities where every neighborhood has its own unique flavor, not at all like the distribution of traits across the city, there is strong sorting of incomes by neighborhood and $I(y;n)$ will be large. How large depends on the relative amount of information needed to describe the system at the local level, Fig. \ref{fig1}B, versus as a whole, Fig. \ref{fig1}C. The mutual information $I(y;n)$ gives a measure of how well a coarse-grained pattern describes a complex system observed at a more disaggregated level. In other words, $I(y;n)$  quantifies the average complexity of any theory of local neighborhood effects versus a theory of the same quantity at the metropolitan level.   

The top and bottom ranked metropolitan areas in the US by the magnitude of $I(y,n)$ are shown in Tables S1-S4. We see that Dallas TX, followed by New York City and New Orleans, LA show the highest $I(y;n)$ in 2010 and that many cities in Texas show in general strong income segregation by neighborhood. This is particularly interesting because these cities are currently among the fastest growing in the nation so that some of the observed income segregation is the result of recent residential choices. Smaller cities, especially in parts of the Midwest (e.g. Wisconsin) but also in other states, show the lowest neighborhood segregation by income.    

\begin{figure}[ht]
\centering
\includegraphics[width=1.0\textwidth]{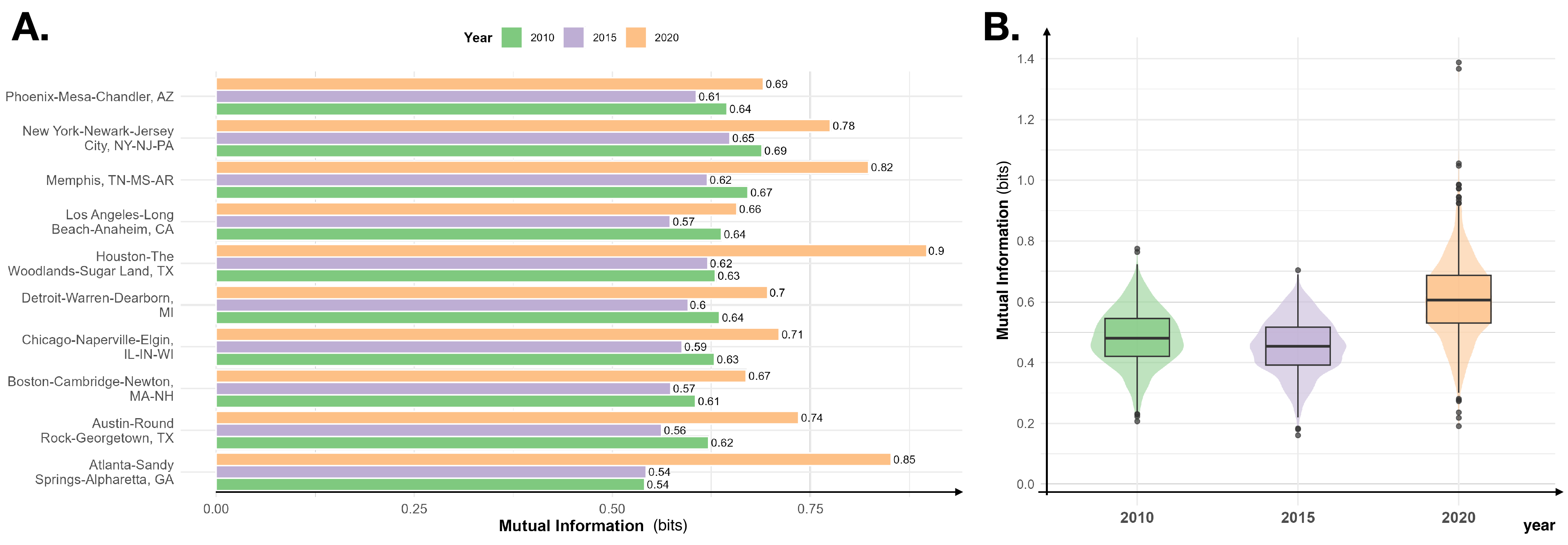}
\caption{Mutual information between income and neighborhoods over time. A. For select large Metropolitan Areas. B. Mutual information across cities over time. The middle line shows the median across cities. Boxes show the inter-quartile range (between the first and third quartiles). Some outliers are shown as points. Shaded areas show the shape of the distribution.  }\label{fig3}
\end{figure}

Figure~\ref{fig3}A shows how the strength of neighborhood effects measured by $I(y;n)$ changes over time, for select larger cities. Note that this analysis is done for three broad temporal periods -- 2010, 2015, and 2020 -- because the data is collected on a 5-year rolling basis. We observe very similar results for 2010 and 2015, but a sharp increase in neighborhood sorting effects in 2020. This effect seems to be more pronounced in some larger cities and is associated with more dispersed observations across all cities, as shown in Fig.~\ref{fig3}B. While we do not have a simple explanation for these observations, the latter part of the 5-year American Community Survey coincided with the COVID-19 pandemic. On the one hand, the pandemic is known to have caused significant non-response bias in the 2020 data in this survey~\cite{Rothbaum2021_ACSNonresponse2020}. On the other hand, there were significant (temporary) population dislocations, with many (more affluent) households leaving central areas of larger cities and creating potentially strong selection effects~\cite{Coven2022UrbanFlight,Foster2024InternalMigrationCOVID,Rebhun2025WhoMovedWhy}.  It will be interesting to continue to monitor this situation to establish the reason and potential reversal of these effects over time.   

\section{Methods}\label{sec11}
\subsection{Data Sources}
Geo-referenced data at the household level (income and population) for the United States is reported at the {\it Census Block Group} level (BK) by the 5-year American Community Survey. Block groups are statistical subdivisions of Census Tracts, which in turn are the basic data collection units for the population census. The boundaries of block groups are generally set so that they contain between 600 and 3,000 people, with a typical size of about 1,500 (or 500 households). Block Groups are spatially contiguous and tile the entire country. Data are aggregated into urban areas defined as Core-Based Statistical Areas (CBSA), which include Micropolitan Statistical Areas and Metropolitan Statistical Areas. Metropolitan areas contain an urban core of 50,000 or more people, while micropolitan areas center around an urban core with 10,000 to 49,999 residents, both defined by adjacent counties with strong economic and commuting ties.(An urban core is a densely settled area.) For simplicity we refer to micropolitan and metropolitan areas together as \textit{Metropolitan Areas}: there are 942 such areas currently in the USA. Counties are the primary legal divisions of States in the U.S., many of which are functioning governmental units whose powers and functions vary from state to state. Counties differ greatly in their areal expansion and populations size. 

\subsection{Units of Analysis: Neighborhood Definitions}

In the main text, we use the term {\it neighborhood} to refer to Census Block Groups, as shown in Figure \ref{fig1}A. Block groups provide an exhaustive tiling of the entire national territory of the United States and its population. In denser areas, Block Groups correspond to smaller land areas, as can be clearly seen in the maps of Fig.\ref{fig1}A. Adopting block groups as proxies for neighborhoods is convenient because they are consistently defined by the US Census Bureau (and similarly by other national census around the world) and provide a universal standard for the study of small area statistics across an entire nation.  For these reasons, they are the most common proxies for social units at this scale (neighborhoods) in the USA. 

A {\it neighborhood} is commonly defined as a geographically localized community where residents share a collective identity, social interactions, and access to local resources. To empirically operationalize this socio-spatial concept, however, a standardized spatial unit is required, and in this study, we follow the common convention of using Census Block Groups as a proxy for neighborhoods (as depicted in Figure 1A). This approach is advantageous because the U.S. Census Bureau intentionally designs block groups to be relatively homogeneous in population and housing, and their exhaustive, standardized coverage of the entire nation enables robust, replicable analysis, making them the default choice for large-scale quantitative research. Nonetheless, this choice has a significant drawback: these administrative boundaries are externally imposed and may not align with residents' perceived community lines or social geographies, potentially bisecting cohesive groups or artificially combining disparate ones. This fundamental tension between administrative convenience and lived experience is why sociologists have long debated the validity of block groups, arguing they can obscure nuanced social dynamics and prompting many local studies to adopt different, more socially meaningful units of analysis (see for example~\cite{hipp_07}).

Our aim here is to demonstrate effects of spatial selection at any given scale. A systematic study of  the strength of spatial selection at different scales using different delineations of neighborhoods is beyond the scope of the present manuscript and will be presented elsewhere.

\subsection{Data limitations}
The American Community Survey (ACS) and the Decennial Census collect household data in small spatial units that allow us to characterize patterns of spatial selection in neighborhoods (i.e., Census Blocks). The ACS is a statistical survey conducted by the US Census Bureau, sent to approximately 250,000 addresses monthly (or about  3 million per year). Unlike the population census (which is strictly a population count), the ACS collects socioeconomic information (for example, on household income). The data are collected primarily by mail, with follow-ups by telephone and personal visits. ACS data are used to make yearly estimates for counties which are then aggregated to provide estimates for States and metropolitan areas.\footnote{For detailed information on the American Community Survey go to \url{www.census.gov/acs/}.} ACS data has an important reporting limitation when it comes to the upper tail of the income distribution: the number of households is listed only for a data bin set by a minimum value ($>\$200$k per household in 2010).

It has been often shown empirically~\cite{Montrol} that, at higher levels of spatial aggregation, the upper tail income distribution deviates from the lognormal pattern reported in Fig. 1C.  Such statistics do, in fact, often follow a Pareto (power-law) distribution for the top richest fraction of 1\%~\cite{Montrol}. Consideration of a finer distribution in this regime is likely to produce even higher atypical values of information for neighborhoods that concentrate such high incomes. In this sense, even though many richer neighborhoods appear the most atypical from the point of view of their income distribution relative to the city at large, Fig. 2B, it is likely that this effect is underestimated as a result of the way data for these incomes are reported.

\subsection{Practical Estimation of Probabilities}
Here, we provide an explicit version of the probability distributions introduced in the main text and the procedure by which they are estimated from discretely binned data.

Let $N$ be the the total number of households in a given city, or the size of that city, for short. Let $N_j$ be the number of households in neighborhood $j$, across all income levels. Then $n_{j,\ell}$ is the number of households in neighborhood $j$, with income (in the interval denoted by) $\ell$.  $N_\ell$ is, correspondingly, the total number of households in the city with income in the interval indexed by $\ell$.  These quantities obey several simple sum rules:
\begin{eqnarray}
&& \sum_j N_j = N, \quad \sum_\ell N_\ell = N, \qquad 
\sum_j n_{j,\ell} = N_\ell, \quad \sum_\ell n_{j,\ell} = N_j.
\end{eqnarray}
Having defined these quantities, which are the ones typically reported by the U.S. Census Bureau, we can provide simple frequency estimators for the several probability densities introduced in the main manuscript. The simplest is $p(n_j)$, the probability of living in a specific neighborhood, which is $p(n_j) = \frac{N_j}{N}$. Analogously, the probability of belonging to a given income level, $\ell$, is $ p(y_\ell) = \frac{N_\ell}{N}.$ The conditional distribution for being in a given neighborhood $j$ given income $\ell$ is $ p(n_j | y_\ell) = \frac{n_{j,\ell}}{N_j}$.
From this and Bayes' relation it follows that
\begin{eqnarray}
p(y_\ell|n_j) = \frac{p(n_j|y_\ell)}{p(n_j)} p(y_\ell) = \frac{n_{j,\ell}}{N_j}.
\end{eqnarray}
The weights $w_{j,\ell}$ are given by
 \begin{eqnarray}
 w_{j,\ell} = N \frac{n_{j,\ell}}{N_\ell N_j}.
 \end{eqnarray}

 Finally, we can check that the properties of the conditional probabilities hold, under these definitions,
 \begin{eqnarray}
 && \sum_j p(n_j|y_\ell) = \sum_j \frac{ n_{j,\ell} }{N_\ell} = \frac{1}{N_\ell} N_\ell = 1. \\
 && \sum_{\ell} p(n_j|y_\ell) N_\ell = \sum_\ell \frac{n_{j,\ell}}{N_\ell} N_\ell = \sum_\ell n_{j,\ell} = N_j. \\
 && \sum_{\ell} p(y_\ell|n_j) = \sum_\ell \frac{ n_{j,\ell} }{N_j} = \frac{1}{N_j} N_j = 1. \\
 && \sum_j p(y_\ell|n_j) N_j = \sum_j \frac{n_{j,\ell}}{N_j} N_j = \sum_j n_{j,\ell} = N_\ell.
 \end{eqnarray}

\section{Discussion}\label{sec12}

 We have shown how intricate local patterns of population traits can develop from broader, more general distributions in larger populations through processes of spatial selection. We also demonstrated that the type and strength of these patterns are most effectively measured in terms of information—a fundamental quantity that provides a unifying language for analyzing complex systems. Cities display simple aggregate regularities alongside rich local variation (often described as  “neighborhoods”). Our central contribution is to show that spatial sorting of households across neighborhoods can be represented, measured, and interpreted as a selection process whose intensity is meaningfully quantified in units of information. 

Selection is a general process by which populations and individuals adapt to their environment by acquiring and processing information and revealing their choices  ~\cite{DonaldsonMatasci2010,hilbert_more_2017,BettencourtGrandisonKemp2025}. Although the word “selection” may evoke biological evolution, here it formalizes a general mechanism by which systems acquire information by differentially amplifying some types over others. As Price noted more than 50 years ago ~\cite{price_selection_1970}, selection is broadly applicable and can be described by the same mathematical formalism. When applied to spatial sorting, this approach allows us to study how local variations can arise within the context of widespread statistical regularities.

In the context of cities, the “types” are income groups, the “environment” is the configuration of housing, amenities, institutions, prices, and social networks, and the amplification arises from households’ choices and constraints interacting with supply, policy, discrimination, and path dependence. The replicator–Bayes equivalence in Eqs.~\ref{eq:1}–\ref{eq:2} makes this explicit: residential choice behaves as inference, where the neighborhoods signal rewards and restrictions, the households update beliefs and options, and the urban system learns by reweighting the joint distribution ($y,n$) towards greater compatibility. 

The quantification of processes of selection in terms of information is still relatively new~\cite{frank_natural_2011,DonaldsonMatasci2010,BettencourtGrandisonKemp2025}. We hope that the results described here provide new additional perspectives into this fundamental identification in respect to the observed multiscale distribution of population and associated features over space in cities. To this end, we have shown how to account for information embedded in the spatial structure of cities at different scales from metropolitan areas to local neighborhoods, thus accounting for the complexity of an overall pattern in terms of both local and global mechanisms. In this way, we hope to bridge two frequently opposing perspectives, by showing how coarse-grained “universality” can co-exist with local choice and specificity. This is particularly poignant for cities, where diversity at the neighborhood levels coexists with relatively simple regularities at the level of functional cities and urban systems~\cite{Sampson2012, Batty2008,Bettencourt2021}. The application of these methods to other locally heterogeneous systems, such as ecosystems or neural networks, is straightforward but requires datasets of comparable scope and quality. 

Models of residential choice have recently been developed representing salient and empirically supported household decisions, characterized by more realistic local contexts, personal traits, and continuous levels of preference~\cite{Bruch2014,BruchMare2006,ReardonBischoff2011}. Estimating these personal and contextual characteristics from empirical data requires a systematic methodology such as the one introduced here, which at once measures the complexity (i.e. the amount of information) necessary at different scales. The recursive property of informational quantities tells us how much of the explanation for an overall pattern may be macroscopic or microscopic by helping us keep track of the information content of models at different scales, Supplementary Text 1.4. Our results show specifically that the aggregate distribution of income for metropolitan New York City is a poor model for most of the city’s neighborhoods, especially its richest and poorest places.

A limitation of the present study is related to the quality of data over time and also to other quantities besides income.  As we have seen for the US, this data is created via a rolling survey, which is sensitive to non-response biases in times of fast change, such during the recent COVID-19 pandemic. In this respect, studies of neighborhood effects related to income using register-based data in nations such as the Netherlands~\cite{Troost2023}, Sweden\cite{Galster2010,Hedman2015}, Norway~\cite{Brattbakk2013,Borgen2025}, or Finland~\cite{Tarkiainen2023,Ristikari2024} present much richer empirical opportunities, also to analyze changes over time and throughout people's life course towards general theory.   

A more systematic understanding of spatial population sorting by personal income and other household characteristics remains at the root of some of the most challenging problems for urban science and policy, including the causes and consequences of economic inequality~\cite{Krivo2013,Chen2012,lens_measuring_2017}, ethnic and racial segregation, disparate access to opportunity~\cite{Chetty2017,chetty_social_2022,lens_measuring_2017} and spatially concentrated (dis)advantage~\cite{Wilson1991,Borgen2025,elliott_effects_1996}, including issues of crime and violence~\cite{Sampson2012,Wilson1991,Intrator2016,Besbris2015}, Supplementary Text 1.5.  Extensions of present models and analytical approaches to more urban systems (nations) and several other demographic dimensions (income, race, education, gender, etc) remain necessary to create better theory of human development and change, and to make urban policy and practice more effective in the face of radically different challenges faced by specific individuals in different places.

\bmhead{Supplementary information}
Includes supplementary text, sixteen figures and four tables.

\bmhead{Acknowledgements}
We thank Christa Brelsford, Elisabeth Bruch, and Laura F\"ursich for discussions. This research was partially supported by the Mansueto Institute for Urban Innovation at the University of Chicago, and by the Arizona State University/Santa Fe Institute Center for Biosocial Complex Systems. 

\section*{Declarations}
\begin{itemize}
\item Funding: This work was partially supported by the Mansueto Institute at the University of Chicago.
\item Competing interests: The authors declare no competing interests.
\item Data, Materials and Code availability; Data, code and materials are available at \url{https://github.com/mansueto-institute/dkl-metric}  
\item Author contribution: LB and JL conceptualized the study and methods, IR performed data curation and computation, LB, JL, JK wrote the paper.
\end{itemize}

\bibliography{sn-bibliography}

\end{document}